\documentclass[12pt,a4paper]{article}
\usepackage{cite}
\usepackage{graphicx, amssymb}
\usepackage{graphics}
\usepackage{amsmath,color}
\usepackage{amsfonts}%
\usepackage{amssymb}
\begin{document}
\begin{center}
\textbf{Numerical Implementation of Generalized Robin--type Wall Functions \\
and Their Application to Impinging Flows}
\end{center}

\begin{center}
\small S.~V.~Utyuzhnikov \\ 
s.utyuzhnikov@manchester.ac.uk \\
School of Mechanical, Aerospace \& Civil Engineering, \\
University of Manchester, \\
PO Box 88, Manchester, M60 1QD, UK
\end{center}

\begin{abstract}
The paper is devoted to the generalized wall functions of Robin--type and their application to near--wall turbulent flows. The wall functions are based on the transfer of a boundary condition from a wall to some intermediate boundary near the wall. The boundary conditions on the intermediate boundary are of Robin--type and represented in a differential form. The wall functions are formulated in an analytical easy--to--implement form, can take into account the source terms of the momentum equation, and do not include free parameters. The log--profile assumption is not used in this approach. A robust numerical algorithm is proposed for implementation of Robin--type wall functions to both finite--difference and finite--volume numerical schemes. The algorithm of implementation of the Robin--type wall functions to existing finite-volume codes is provided. The axisymmetric impinging jet problem is numerically investigated for different regimes on the base of the wall-functions implemented to the high-Reynolds-number $k- \epsilon$ model.
\end{abstract}

\begin{flushleft}
\bfseries 1 Introduction
\end{flushleft}
Problems related with turbulent near wall flows appear in many industrial applications. It is well-knoen that turbulence vanishes near a wall due to both the no-slip boundary condition for the velocity and the blocking effect caused by the wall. In the vicinity of the wall, there is a thin sublayer with predominantly molecular diffusion and viscous dissipation. The sublayer has a substantial influence upon the remaining part of the flow. An adequate resolution of a solution in the sublayer requires a very fine mesh because of the thinness of the sublayer and high gradients of the solution. It makes the model to be time consuming and often it is not suitable for real design. Because of the low turbulent Reynolds number in the sublayer, the models that resolve the sublayer are called low-Reynolds-number (LR) models. 

In turn, the high-Reynolds-number (HR) models do not resolve the viscous sublayer. It significantly saves computational efforts  \cite{wall-function, development}. In the HR models, the boundary conditions or near-wall profiles are represented by wall functions. The wall functions usually are semi-empirical and have very limited applications \cite{wall-function, development, progress, heat, numerical}.  First wall functions are based on the log-law profile assumption for the velocity \cite{heat, numerical}. In addition, their formal extension to complex flows demands time-consuming iterations for calculating the skin friction included in the log-law profile. A substantial disadvantage of these wall functions is a strong dependence on the near wall mesh used. This problem is especially pronounced if the first mesh point is located inside the viscous sublayer. To avoid this, the scalable wall functions are suggested in \cite{Esch}. Wilcox assumes \cite{Wilcox} that the pressure gradient must be taken into account to avoid the mesh dependence. Yet, the recently suggested adaptive wall functions \cite{Kalitzin} overcome this limitation by using look-up tables for turbulent quantatives and skin friction.

In more recent wall functions \cite{wall-function, development, progress, strategy, Kidger} source terms, such as the pressure gradient, might be taken into account. The numerical comparisons done in \cite{wall-function, development, progress, Kidger} showed that such advanced wall functions give substantially better prediction than the standard wall functions. In \cite{progress, Kidger}, the analytical wall functions are obtained by approximate integrating boundary-layer-type equations in the wall vicinity using the assumption that all terms besides the diffusive one are constant. At the wall, the boundary conditions are the same as those used in the LR models. An analytical profile for the turbulent viscosity are then used in the cell nearest to the wall to reconstruct the near-wall solution. The wall functions for the turbulent kinetic energy and its dissipation are based on the local analytical solution for the velocity in the near-wall cell. In computations, the numerical flux to the wall is taken from the previous iteration as it is performed in the case of the standard wall functions. Although approaches \cite{wall-function, development, progress, strategy, Kidger} allow one to make substantially better prediction in comparison to the standard methods, their realization seems to be quite complicated. The wall functions \cite{wall-function, development, progress, strategy, Kidger} are only represented in a finite-difference form. Although this form is suitable for finite-volume algorithms, its use for finite-difference approximations is not clear. Similar to the standard wall functions, this approach faces substantial problems if the nearest to the wall cell is in the viscous sublayer. Also, it is important to note that the second to the wall cell cannot be much smaller or bigger than the first one because of the integration over the first cell. 

The method of boundary condition transfer is suggested in  \cite{Ut, Ut2}. The method allows us to transfer a boundary condition from the wall to some intermediate surface. The boundary condition is transfered either approximately (analytically) or exactly (numerically). The boundary conditions on the intermediate surface are always of Robin--type (or mixed type) and represented in a differential, mesh independent, form. These boundary conditions are set on both a function and its normal derivative. Therefore, their realization does not require additional iterations between, for example, the velocity and skin friction. This brings an additional robustness to the algorithm of their implementation. These boundary conditions are interpreted as generalized (Robin--type) wall functions. Another advantage of these wall functions is related with their universal formulation for all dependent variables. 

The Robin--type wall functions take into account the influence of the source terms in governing equations. The location of the point, to which the boundary conditions are transferred, does not make any considerable effect on the mesh distribution nearby this point. The wall functions can be implemented in both finite-difference and finite-volume approximations. It is shown how the wall functions can be implemented in existing codes. Preliminary tests for channel flow \cite{UtNM} and impinging jet \cite{Ut2} have shown promising results in terms of both accuracy and efficiency gains. In comparison to the analytical wall functions \cite{progress, Kidger}, the key advantages of the Robin--type wall functions are related with their robust implementation and universal differential formulation.   

In the paper below the Robin--type wall functions are implemented in the $k - \epsilon$ model and applied for the axisymmetric impinging jet problem. The computational results are compared against available experimental data. Implementation of the wall functions to both finite-volume and finite-difference schemes are discussed.

\begin{flushleft}
\bfseries 2 Generalized (Robin--type) wall functions
\end{flushleft}
In order to formulate the Robin--type wall functions, first let us record the governing equations in the following general form:

\begin{equation} \label{E:muy=R}
  \left(\mu u_y\right)_y=R_h
\end{equation}
with Dirichlet boundary condition on the left-hand side: 

\begin{equation}
	u\left(0\right)=u_0
\end{equation}

Equation \eqref{E:muy=R} represents the general form of the boundary-layer-type equation. The right-hand side $R_h$ is an appropriate source term including, e.g., the pressure gradient in the momentum equation. 

If the right-hand side $R_h = const$, then the Robin--type wall functions are formulated as follows \cite{Ut, Ut2}:

\begin{equation} \label{E:ustar}
  u(y^*)=u_0 + f_1 \frac{du}{dy}(y^*) - \frac{R_h}{\mu(y^*)} f_2,
\end{equation}
where
\begin{equation} \label{E:f1}
	f_1 = \int^{y^*}_{0}\frac{\mu(y^*)}{\mu(y)}dy, 
 ~f_2 = \int^{y^*}_{0}\frac{\mu(y^*)}{\mu(y)}(y^*-y)dy.
\end{equation}

Relation \eqref{E:ustar} can be interpreted as a boundary condition of Robin--type transferred from a wall ($y = 0$) to some point $y^*$. This boundary condition can be either exact (if the exact function of $\mu$  is used in \eqref{E:f1}) or approximate (if $\mu$  is estimated by one way or another). One should emphasize that the Robin--type boundary condition is set for both a function and its derivative. It is easy to see that the mesh distribution nearby the point $y^*$ can be independently chosen on the location of this point. Implementation of Robin--type conditions to both finite--difference and finite--volume schemes are considered below. 

In the general case $R_h = R_h (y)$, we have
\begin{equation} \label{E:ustarfl}
  u(y^*)=u_0 + f_1 \frac{du}{dy}(y^*) - \left(\int^{y^*}_{0}R_hdy\right)\frac{f_2}{y^* \mu(y^*)},
\end{equation}
where
\begin{equation} \label{E:f1fl}
	f_1 = \int^{y^*}_{0}\frac{\mu(y^*)}{\mu(y)}dy, 
 ~f_2 = y^*\int^{y^*}_{0}\frac{\mu(y^*)}{\mu(y)}\left(1-\frac{\int^{y}_{0}R_hdy}
  {\int^{y^*}_{0}R_hdy}\right)dy
\end{equation}

Having assumed that the coefficient   varies piece--wise linearly 
\[
\mu=
  \begin{cases}
   \mu_w,                                             &\text{if $0  \leq y\leq y_v$}\\ 
   \mu_w + (\mu^* - \mu_w)\frac{y-y_v}{y^*-y_v},      &\text{if $y_v\leq y\leq y^*$},
  \end{cases}
\]
it is possible to obtain analytical expressions for $f_1$ and $f_2$ if $R_h = const$ and 
$y_v \leq y^*$ :
\begin{equation} \label{E:forml}
	f_1 = \alpha_\mu y_v(1 + \theta\ln\alpha_\mu),
 ~f_2 = \alpha_\mu y_v\left[(1-\theta)y^* +  
        y_v(\theta^2\alpha_\mu\ln\alpha_\mu-1/2+\theta)\right],
\end{equation}
where $\alpha_\mu = \mu^*/\mu_w, \theta^{-1} =\frac{\mu^*-\mu_w}{\mu_w}\frac{y_v}{y^*-y_v}$. The parameter  $\theta$ represents cotangent of the inclination angle of the dependence  $\mu/\mu_w$ on $y/y_v$. 

If $R_h = const$, the wall--flux can be found as follows \cite{Ut2}:

\begin{equation} \label{E:frict}
  \tau_w=\frac{\mu^*u(y^*)}{f_1}+\left(f_2/f_1-y^*\right)R_h.  
\end{equation}

This formula can be easily generalized on the case of a variable right-hand side $R_h$ using \eqref{E:ustarfl}.

The method of boundary condition transfer technique can be used to derive the wall functions for the tangential and normal velocity components $U$ and $V$, temperature $T$ , and turbulent kinetic energy $k$. 

Having neglected diffusion parallel to the wall, the momentum and enthalpy transport equations can be written in the Cartesian coordinate system $(x, y)$ as follows: 

\begin{align}
  \label{E:rmeq}
	\frac{\partial}{\partial y}\left[(\mu_l + \mu_t) \frac{\partial U}{\partial y} \right] &= 
	\rho U\frac{\partial U}{\partial x} + \rho V\frac{\partial U}{\partial y} + 
	\frac{\partial P}{\partial x} \\
	\frac{\partial}{\partial y}\left[(\mu_l + \mu_t) \frac{\partial V}{\partial y} \right] &= 
	\rho U\frac{\partial V}{\partial x} + \rho V\frac{\partial V}{\partial y} + 
	\frac{\partial P}{\partial y} \\
	\frac{\partial}{\partial y}\left[(\frac{\mu_l}{Pr} + \frac{\mu_t}{Pr_t}) 
	\frac{\partial T}{\partial y} \right] &= 
	\rho U\frac{\partial T}{\partial x} + \rho V\frac{\partial T}{\partial y} 
\end{align}
Here $\mu_l$ and $\mu_t$ are the laminar and turbulent viscosities, accordingly; $Pr$ and $Pr_t$ are Prandtl numbers; $U$ and $V$ are the velocity component in the $(x, y)$ coordinate system; $\rho$  is the density; $P$ is the pressure.

The intermediate boundary conditions for $U$, $V$ and $T$ at point $y^*$ are given by \eqref{E:ustarfl} upon substitution $U$, $V$ or $T$ instead of $u$ accordingly. Evidently, the coefficient $\mu$  in \eqref{E:muy=R} must be considered as either  $\mu_l  +  \mu_t$ or $ \mu_l/Pr  +  \mu_t/Pr_t$. In the case of the momentum equation $u_0 = 0$. If $y^*$ is chosen in the vicinity of the wall, the right-hand side $R_h$ can be simply evaluated at $y^*$. Thus, in the case of the momentum equations and enthalpy the relative right-hand sides are as follows:

\begin{eqnarray}\label{E:RhU}
	R_h &=& R_{hu}\equiv\rho \left(U\frac{\partial U}{\partial x}(y^*)
	+V\frac{\partial U}{\partial y}(y^*)\right)+\frac{\partial P}{\partial x}(y^*),
	\\
	R_h &=& R_{hv}\equiv\rho 
	\left(U\frac{\partial V}{\partial x}(y^*)
	+V\frac{\partial V}{\partial y}(y^*)\right)+\frac{\partial P}{\partial y}(y^*),
	\\
	R_h &=& R_{ht}\equiv\rho (U\frac{\partial T}{\partial x}(y^*)+V\frac{\partial T}
	{\partial y}(y^*))\\	
	\nonumber
\end{eqnarray}
Thus, all the terms of the parabolized (reduced) Navier-Stokes equations (PNS) \cite{Rub} are taken into account. It worth noting that in the boundary condition \eqref{E:ustar} for the normal velocity it is not assumed to be zero. Thus, these wall functions do not have direct restrictions to their exploration in modeling separated flows.  

Unlike \cite{progress}, a similar approach is applied to the equation for the turbulence kinetic energy as well:
\begin{equation}\label{E:kfull}
	\frac{\partial}{\partial y}\left[(\mu_l + \frac{\mu_t}{Pr_k}) \frac{\partial k}{\partial y}
	\right] = \rho U\frac{\partial k}{\partial x} + \rho V\frac{\partial k}{\partial y} - P_k +
	\rho\epsilon,
\end{equation}
where $P_k$ is the production of the turbulent kinetic energy, $\epsilon$  is its dissipation; $Pr_k$ is the Prandtl number. 

Having evaluated the convective terms, we obtain the following expression for the right-hand side $R_h$:
\begin{equation}\label{E:Rh}
	R_h = R_{hk}\equiv\rho \left(U\frac{dk}{dx}(y^*)+V\frac{dk}{dy}(y^*)\right)+ \rho\epsilon - \mu_t \left(\frac{dU}{dy}\right)^2
\end{equation}
Having assuming a piece-wise linear behavior of the function $\mu_t$:
\begin{equation}\label{E:mu}
	\mu_t=
  \begin{cases}
   0,                                             &\text{if $0  \leq y\leq y_v$}\\ 
   \mu^{*}_{t}\frac{y-y_v}{y^*-y_v},               &\text{if $y_v\leq y\leq y^*$},
  \end{cases}
\end{equation}
where $y_v$ is the thickness of the viscous sublayer near the wall, the coefficients $f_1$ and $f_2$ in \eqref{E:f1fl} (the latter term only if $R_h = const$) can be evaluated by \eqref{E:forml}.

For the momentum equation
\begin{equation}\label{E:alfengy}
	\alpha_\mu = \mu^*/\mu_l, ~\theta = \frac{y^* - y_v}{y_v}\frac{\mu_l}{\mu^{*}_{t}}, 
 ~\mu^{*} = \mu_l + \mu^{*}_{t},
\end{equation}
while in the case of the energy equation
\begin{equation}
	\alpha_\mu =\frac{\mu_l/Pr + \mu^{*}_{t}/Pr_t}{\mu_l/Pr}, 
	~\theta = \frac{y^* - y_v}{y_v}\frac{Pr_t}{Pr}\frac{\mu_l}{\mu^{*}_{t}}
\end{equation}

If the turbulent viscosity  $\mu^{*}_{t}$ in \eqref{E:mu} is evaluated as follows \cite{progress}:
\begin{equation}\label{E:mut1}
	\mu^{*}_{t} = C_\mu C_l \rho \frac{\sqrt{k^*}}{\mu_l}y_v\frac{y^*-y_v}{y_v}\mu_l=
	C_\mu C_l Re_v\frac{y^*-y_v}{y_v}\mu_l\approx2.5\frac{y^*-y_v}{y_v}\mu_l,
\end{equation}
where $k^* = k(y^*), C  = 0.09, C_l = 2.55, Re_v\equiv \frac{\rho \sqrt{k^*}y_v}{\mu_l}=10.8$,   then $\theta$   is a constant equaled to $0.4$ for the momentum equation.

It has been found from the computations that it is more accurate to evaluate the turbulent viscosity at $y^*$ from the HR $k-\epsilon$ model directly
\begin{equation}\label{E:mutstr}
	\mu_t = C_\mu \rho \left(k^*)\right)^2/\epsilon
\end{equation}
rather than from equation \eqref{E:mut1}.

The sublayer thickness $y_v$ is evaluated as follows \cite{progress}: 
\begin{equation}\label{E:yv}
	y_v = Re_v \mu_l/\left(\rho \sqrt{k_v}\right),
\end{equation}\label{E:Re_v}
where $k_v$ is the value of $k$ at the edge of the viscous sublayer. 

If $y^* < y_v$, then the boundary conditions are set inside the sublayer, and formulas \eqref{E:forml} are not formally valid. It is suggested to pose the boundary conditions at the edge of the sublayer $y = y_v$ as in \cite{Esch} because $y_v$ is small enough. Then, the coefficients $f_1$ and $f_2$  in \eqref{E:ustarfl} can be evaluated as follows: 
\begin{equation}
	f_1 = \alpha_\mu y_v, \quad f_2 = \alpha_\mu y^{2}_{v}/2.
\end{equation}

It is then assumed that the turbulent viscosity $\mu_t$ reaches value \eqref{E:mutstr} at the edge of the viscous sublayer immediately. These boundary conditions are consistent with boundary conditions \eqref{E:forml} taking in the limit  $y^* \rightarrow y_v$ or $\theta \rightarrow 0$.

The dissipation of the turbulent kinetic energy  $\epsilon$ is evaluated as in \cite{progress}:
\begin{equation}\label{E:epsilon}
	\epsilon (y) = 
  \begin{cases}
   \frac{\left(k^*\right)^{3/2}}{C_l y_d},     &\text{if $y < y_d,$}\\
   \frac{\left(k^*\right)^{3/2}}{C_l y},       &\text{else},
  \end{cases}
\end{equation}
where $y_d = 2C_l \mu_l/\left(\rho \sqrt{k^*}\right)$.  

The wall function for the turbulent energy $k$ is used in form \eqref{E:ustarfl}, \eqref{E:f1fl} and depends on the right-hand side $R_{hk}(y)$ represented by equality \eqref{E:Rh}. It includes the dissipation $\epsilon$  and derivative $dU/dy$. The former term is taken from \eqref{E:epsilon} while the latter term can be evaluated in the interval $[0, y^*]$ from the reduced momentum equation \eqref{E:rmeq}, \eqref{E:RhU}:
\begin{equation}\label{E:mutl}
	(\mu_l + \mu_t)dU/dy = \left[(\mu_l + \mu^{*}_{t})U(y^*) +
	 f_2R_{hu}\right]/f_1+(y-y^*)R_{hu},
\end{equation}
where the turbulent viscosity $\mu_t$ is defined by \eqref{E:mu}. Equation \eqref{E:mutl} is obtained by the integration of equation \eqref{E:rmeq} and use relation \eqref{E:ustar} to exclude $dU/dy(y^*)$. Thus, this equation relies on the PNS assumptions used. 

To evaluate $y_v$ from \eqref{E:yv}, it is possible to use the value $k^*$ instead of $k_v$. It allows us to simplify the evaluation algorithm for $y_v$ substantially. First, a similar opportunity was noticed in \cite{progress}. The estimation of $ Re_v $ is varied between 10.8 and 20 \cite{progress, Lars}. It corresponds to the interval between the upper limit of the viscous sublayer and the point at which the linear and logarithmic parts of the velocity profile intersect for the channel flow \cite{Moser}. It is not clear which value in this interval is most appropriate to approximation \eqref{E:mu}. In all computational results given below, $ Re_v = 12$. 
  
It worth noting that the coefficients $f_1$ and $f_2$ in wall functions \eqref{E:ustar}--\eqref{E:forml} depend only on $y^*$ and $k^*$. The latter value is determined from the solution of the HR model at the boundary point $y^*$. Hence, the intermediate boundary conditions \eqref{E:ustarfl} at $y = y^*$ complete the boundary-value problem in the interval $[y^*, y_e]$ and can be considered as generalized wall functions. These boundary conditions are of Robin--type and similar to the "slip boundary condition" at the edge of the Knudsen--layer in aerodynamics. One should note that the boundary conditions are linear with respect to the leading variable. As it follows from \eqref{E:ustar} and \eqref{E:ustarfl}, the source terms in the wall functions can only be essential far enough from the wall because of the quadratic dependence of $f_2$ on $y^*$.

The HR solution obtained in $[y^*, ~y_e]$ can be extended to interval $[y_v, ~y^*]$ using the analytical solution in this interval:
\begin{eqnarray}\label{E:upart}
	u(y) &=& u(0) + \phi_1(y) u^{*}_{y}-\phi_2(y)\frac{R_h}{\mu^*}, \\
\nonumber 
  \phi_1 &=& \alpha_\mu y_v \left(1+\theta \ln \Omega (y)\right),\\
\nonumber 
  \phi_2 &=& \alpha_\mu y_v \left[y^* - \theta y + (\theta^2 \alpha_\mu -1/2+\theta)y_v\right],\\
\nonumber 
  \Omega &=& 1 + (\alpha_\mu -1)\frac{y-y_v}{y^*-y_v}.
\end{eqnarray}

Thus, the intermediate boundary is not necessarily to be related to the nearest to the wall cell. It is possible to take $y^*$ far enough from the wall and complement the solution on the region of the sublayer by \eqref{E:upart}. 

It worth noting that, although the problem is solved in the bulk domain $[y^*,~y_e]$ only, the flux to the wall (e.g., skin friction) can be easily evaluated considering \eqref{E:frict} (or its analogy for the temperature in the case of heat flux). 

Thus, the developed wall functions can be applied to all dependent variables but $\epsilon$ in a uniform manner. It is possible to show that the coefficients $f_1$ and $f_2$ can be determined analytically even in the case of the turbulent kinetic energy $k$. This fact can be useful for saving computer resources. For engineering purposes it worth considering a simplified version of the wall functions corresponding to $f_2 = 0$. Since the coefficient $f_1$ can be chosen to be the same for all variables, the wall functions become fully identical in this case. 

The generalized Robin--type wall functions are not based on a numerical approximation in the inner region $[0, y^*]$, as in \cite{wall-function, development, progress, Kidger}, therefore the location of the intermediate boundary is not very substantial for the mesh distribution in the bulk domain. It means we can choose, e.g., a fine mesh despite a relatively big value of $y^*$  (or vice versa) without loose of stability.  

In this paper we make the main focus on the analytical evaluation of the coefficients $f_1$ and $f_2$ in \eqref{E:ustar}. Yet, the coefficients can be evaluated numerically by integrating LR equations in the interval $0, ~y^*$. It may lead to the decomposition method \cite{Ut, Ut2}. Also, this approach can be naturally integrated in the numerical wall-functions \cite{development, strategy}. 

\begin{flushleft}
\bfseries 3 Numerical implementation of Robin--type wall functions
\end{flushleft}
The Robin--type wall functions can be implemented to both finite--difference and finite--volume RANS approximations. In this section, some aspects of their robust implementation are considered. 

A general remark related with robust implementation is as follows. Boundary conditions of Robin--type are set on both a function and its derivative. Upon approximating the derivative, both terms should be considered at the same iteration (or time step). Taking into account one of the terms from a previous iteration leads to additional iterations, at least. It is easy to see this property in the case of a linear equation. Only simultaneous consideration of both terms provides an iterationless solution.

The boundary condition is represented by \eqref{E:ustarfl} for any $y^*$ in the vicinity of the wall even if $y^*$ vanishes. This boundary condition fully replaces the original boundary condition $u(0) = u_0$. 

In numerical simulation of turbulence, the finite--difference numerical schemes preserving positiveness of a solution \cite{positiv} are very efficient because unknown variables such as the turbulent kinetic energy $k$ or its dissipation $\epsilon$  must be positive. The following numerical procedure can be used for developing the positive definite schemes in solving boundary-value problems with Robin--type boundary conditions \cite{Ut2}.

Boundary condition \eqref{E:ustarfl} can be rewritten in the following general form:
\begin{equation}\label{E:k0}
	k(0) = \alpha dk/dy(0) + \beta,
\end{equation}
assuming that both the function $k$ and its derivative $dk/dy$ are positive. This assumption is valid in the case of real physical problems for the turbulent kinetic energy in the wall vicinity. The coefficient $\alpha$  is positive because $f_1$ is always positive but the coefficient $\beta$  can be negative (mostly, where $\epsilon > P_k$). In computations it can lead to a negative value of $k$. To avoid such a case, it is suggested to rewrite \eqref{E:k0} in the following form if $\beta  < 0$:
\[	k(0) = \alpha dk/dy(0) + \beta \frac{k(0)}{k^-(0)},  
\]
or
\begin{equation}\label{E:k01}
		k(0) = \tilde{\alpha} dk/dy(0),
\end{equation}
where $\tilde{\alpha} = \frac{\alpha}{1-\beta/k^-(0)}$  and $k^-(0)$  is the value of $k(0)$ taken from the previous either time step or iteration. 

At last one should note that at some distance from the wall the derivative $dk/dy$ becomes negative. Yet, in this area the function $\beta$ is positive $(P_k > \epsilon)$ and, therefore, there are no principal difficulties in reaching a positive solution.      

Though Robin--type boundary condition \eqref{E:ustarfl} can be set at the wall, its implementation to existing codes based on finite--volume schemes is more easy in the following treatment. Assume that unknown variables are defined at the centers of cells. For the sake of simplicity let us consider 1D approximation in the normal to the wall direction. It is enough to consider approximation at the nearest to the wall cell since the rest approximation is remained without any modification. Let us denote values at the centre of the cell by index 1/2 and values at the edge, opposite to the wall, by index 1. Then, considering $y^* = y_{1/2}$ we have a relation in the following form:

\begin{equation}\label{E:u1/2}
		u_{1/2} = u_w + f_1 \frac{du}{dy}_{|1/2} + \tilde{f}_2,
\end{equation}
where $\tilde{f}_2=- \frac{\int^{y_{1/2}}_{0}R_hdy}{y_{1/2} \mu(y_{1/2})}f_2$. 
At the first cell the governing equation \eqref{E:muy=R} is then integrated only from $y_{1/2}$ to $y_1$:
\begin{equation}\label{E:u01}
		\mu\frac{du}{dy}_{|1} - \mu\frac{du}{dy}_{|1/2}u_{1/2} = \int^{y_1}_{y_{1/2}}R_hdy
\end{equation}
Equation \eqref{E:u01} represents the approximation of the governing equation in the near-wall cell. Here, the flux $\mu\frac{du}{dy}_{|1}$ is approximated by an ordinary approach while the flux at $y_{1/2}$ is obtained from \eqref{E:u1/2}.
A typical approximation can be written as follows:
\begin{equation}\label{typapr}
		\mu_1\frac{u_{3/2}-u_{1/2}}{y_1}-\mu_{1/2}\frac{u_{1/2}-u_w-\tilde{f}_2}{f_1} = \int^{y^1}_{y_{1/2}}R_hdy
\end{equation}
This kind of approximation is used for all variables, but $\epsilon$, including $k$. In the last case, the right-hand side is rapidly changed and the integral in the right-hand side must be evaluated accurately enough. It can be easily done either numerically or analytically using the analytical expression for the velocity gradient \eqref{E:mutl}. 

If $y^* > y_v$, the considered above approximation is close to the approximation based on the analytical wall functions (AWF) \cite{progress}. The principal difference is related with robustness. In the case of the Robin--type wall functions both the function and its derivative (flux) are simultaneously taken into account while in realization of the AWF the flux is calculated in iterations. This difference becomes more clear in the case of a linear equation. Then, the Robin--type formulation of the boundary conditions does not require any iterations. 

At last, a brief comment can be done with regard to a staggered mesh. In this approach the velocity is defined at the vertexes of a cell. In our consideration this means $u_1$ is known instead of $u_{1/2}$. In this case, the simplest way to remain the uniform approach is based on the Taylor expansion:
\begin{equation}\label{Taylor}
	u_1 = u_{1/2} + y_{1/2}\frac{du}{dy}_{|1/2} + \frac{y^2_{1/2}}{2}\frac{d^2u}{dy^2}_{|1/2}.
\end{equation}
This yields a required relation between $u_{1/2}$ and $u_1$ since the first derivative takes place in \eqref{E:u1/2} and the second derivative can be evaluated via the right-hand side $R_h$. Thus, the flux at the nearest to the wall cell is as follows:

\begin{equation}
  \mu_{1/2}\frac{du}{dy}_{|1/2}=\mu_{1/2}\frac{u_1-u_w}{f_1+h/2}-R_h\frac{h^2/8-\tilde{f}_2}{f_1+h/2}.
\end{equation}

The same technique with slight modifications can be used for implementation to unstructured codes.  

\begin{flushleft}
\bfseries 4 Impinging jet
\end{flushleft}
The problem of impinging jet appears in many industrial applications related with either heating or cooling processes. The heat transfer problem on a turbulent circular jet impinging onto a flat is well studied experimentally \cite{Cooper, Baughn, Baughn1, Yan}. Therefore, it became a widespread test case for different turbulent models including the LR \cite{ Graham, Sunden, Pollard, Yoon, Park, Dick} and HR $k-\epsilon$ models \cite{strategy, Esch, Pollard, A-F}. 

The problem is formulated as follows. A fully turbulent air jet, generated in a pipelike nozzle, impinges on a flat surface at the right angle. The Reynolds number, based on the nozzle diameter $D$ and the bulk velocity, equals to $Re = 23000$ and $Re = 70000$. The distance $L$ between the nozzle and the surface is varied between $2D$ and $14D$. Air is treated as an ideal gas and considered under normal conditions at temperature $293 K$. The heated surface has constant temperature $T_w=314.9 K$.   

The computational domain spans $13 D$ in the radial direction. The grid includes 150x100 (axial x radial) nodes and 150x200 nodes. For the validation purposes, preliminary comparisons of the results obtained on different meshes were done to check grid sensitivity. The boundary conditions at the edge of the nozzle are specified using the profiles for a fully developed turbulent pipe flow. The computations of the local Nusselt number are done for the different values of $y^*$ or $Re_{y^*}\equiv \rho\sqrt{k^*}y^*/\mu_l$ calculated at the stagnation point. 

Linear eddy-viscosity model (EVM) drastically overpredict the turbulent kinetic energy in the stagnation point region by an order of magnitude \cite{A-F}. It inevitably leads to the considerable overestimation of the heat flux. As a result, the linear LR $k-\epsilon$ models give unacceptable overprediction by a factor of two, even more \cite {development, Park, Dick, Sunden}. Furthermore, the linear $k-\epsilon $ model, as well as other EVM, is not entirely justified around the stagnation point because of the anisotropy of the flow. To improve prediction, along with the non-linear EVM, some modifications of the EVM are used including the implementation of a realizability constraint \cite{Sunden}, introduction of empirical formulas for the Prandtl number \cite{Park} and heat flux \cite{Esch}. The application of the wall functions, in most cases, is also showed a poor performance \cite{development, Pollard, A-F}. More or less reasonable prediction was achieved in \cite{development} using the Chieng -- Launder wall function \cite{Chieng} and the scalable wall functions \cite{Esch}. It is to be noted that in the latter case the empirical correlation was used for the local heat flux. The generalized wall functions were applied in \cite{Ut2} to simulation of the impinging jet at $Re = 23000$ with $L/D = 2$ and $L/D = 6$. Apart from the heat flux, the prediction of the wall friction is considered in \cite{Ut2}. As was noted above, the overprediction of the heat flux was obtained in the computational solution. 

\begin{figure}[tbp]
\centering
\scalebox{.49}{
\includegraphics{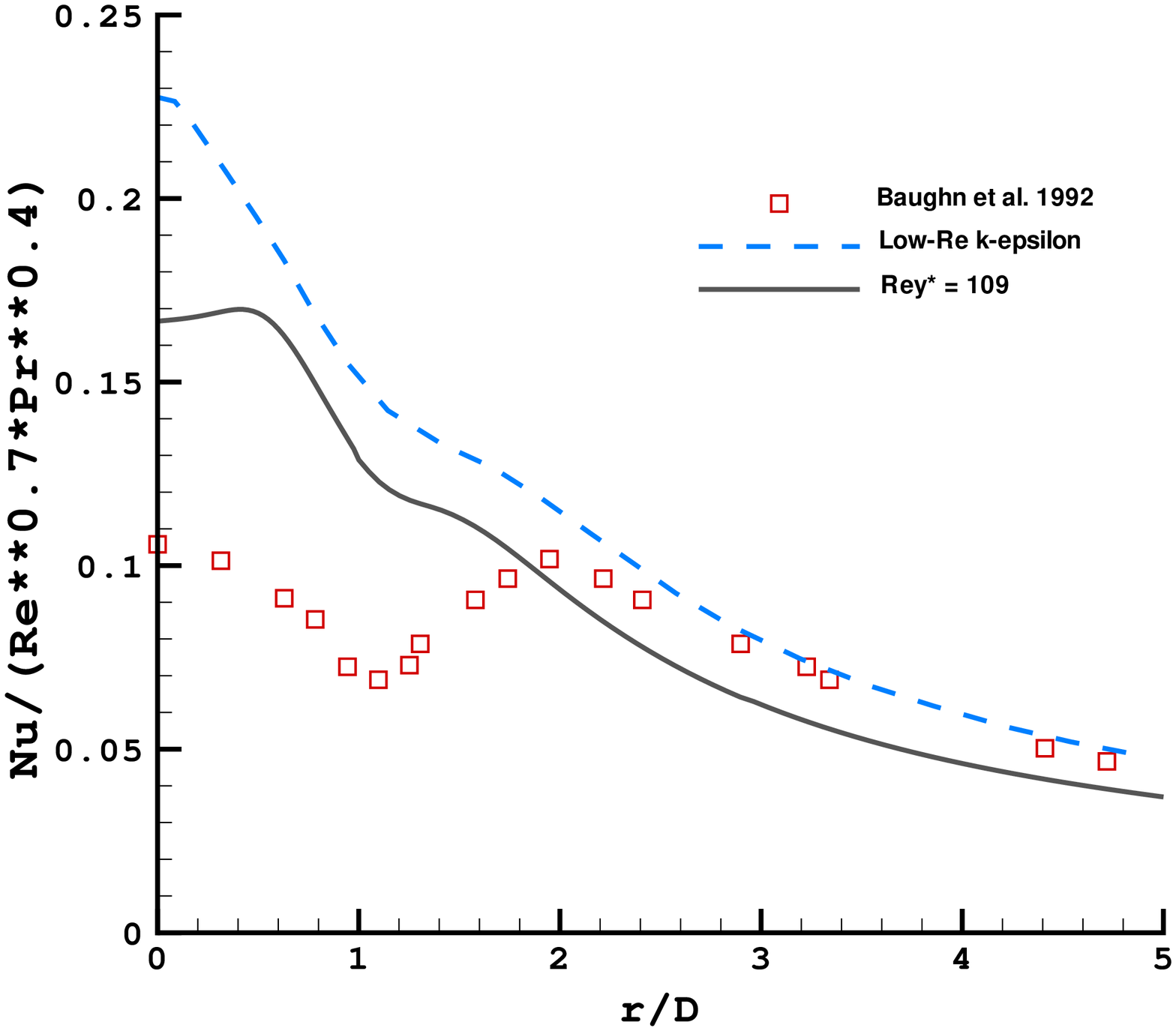}
}
\caption{Local Nusselt number for the impinging jet. Comparison between HR, LR solutions and  experiment for $Re = 70000$ and $L/D=2$.}
\label{Fig:nusselt2}
\end{figure}

\begin{figure}[tbp]
\centering
\scalebox{.49}{
\includegraphics{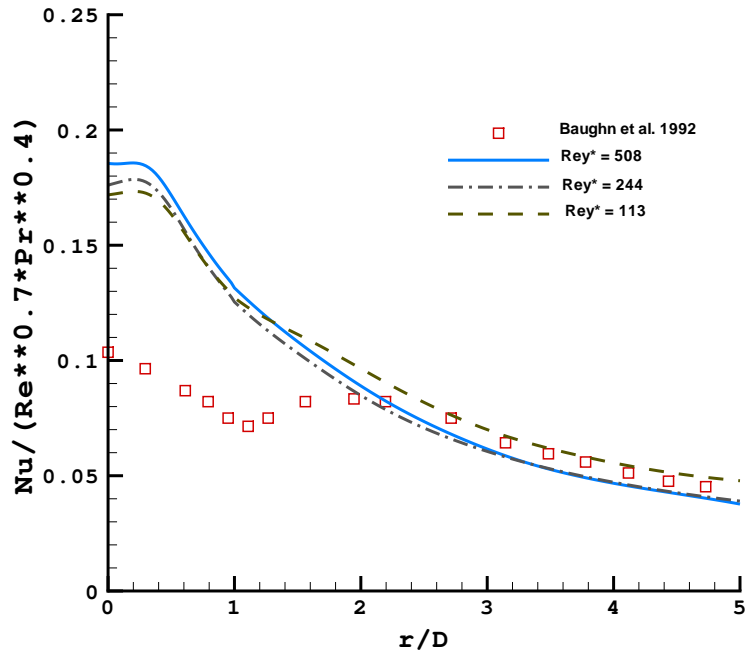}
}
\caption{Local Nusselt number for the impinging jet. Comparison of computational solution  for different $y^*$ against  experiment for $Re = 70000$ and $L/D=4$.}
\label{Fig:nusselt4}
\end{figure}

The same effect but more expressive is observed at $Re = 70000$. In Fig.\ref{Fig:nusselt2}, the computational results are compared against the experimental data for $L/D = 2$. Here and below, the local Nusselt number is scaled by $Re^{0.7}Pr^{0.4}$ where $Pr = 0.9$. The solution \cite{Graham} based on the low-Reynolds number $k-\epsilon$ model predicts a substantially higher heat flux than the high-Reynolds number model. This effect was obtained in many other publications including cited above. The solution based on the wall functions corresponds to $Re_{y^*} = 109$. It is to be noted that the dependence of the solution on the parameter $y^*$ is quite weak. In the next example with $L/D = 4$, the solutions corresponding to different values of $Re_{y^*}$ are shown in Fig.\ref{Fig:nusselt4}. Though the value of $y^*$ is varied by an order of magnitude, the curves are quite close each other. 

\begin{figure}[tbp]
\centering
\scalebox{.49}{
\includegraphics{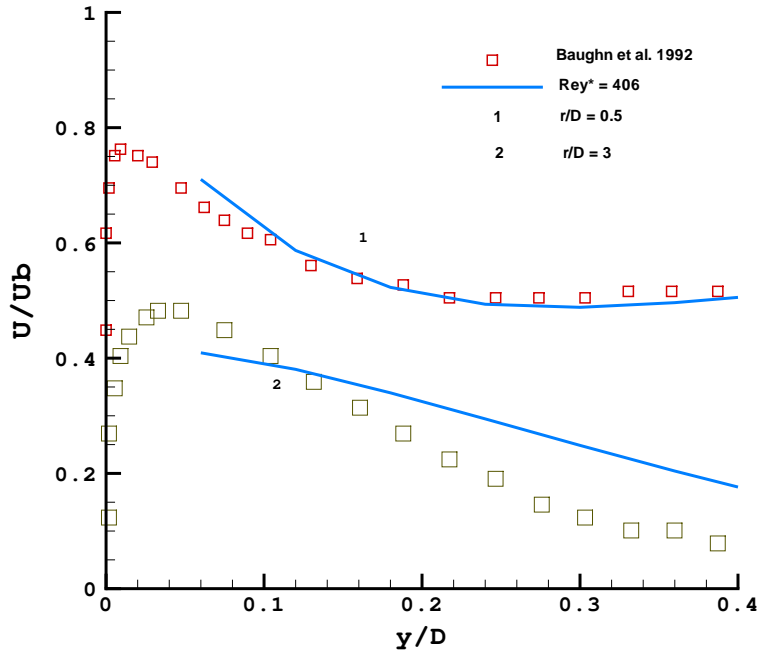}
}
\caption{Mean velocity in the impinging jet at $r/D = 0.5; 3$. Comparison of computational solution against  experimental data for $Re = 70000$ and $L/D=6$.}
\label{Fig:velimpj2}
\end{figure}

The mean velocity profiles divided by the bulk velocity are shown in Fig.\ref{Fig:velimpj2} for $L/D=6$.  The experimental data are represented by square symbols while the computational results are shown by the curves. At the region of the low mean velocity nearby the axis of symmetry ($r/D=0.5$) the prediction of the velocity is quite accurate. At $r/D = 3$, where the flow is decelerated, the prediction is not so good.  At this location, substantial underprediction  of the velocity in the near wall region and overprediction in the outer region were earlier  noted for both the LR and HR linear $k-\epsilon$ models \cite{Esch, Graham, Dick}.  

\begin{figure}[tbp]
\centering
\scalebox{.49}{
\includegraphics{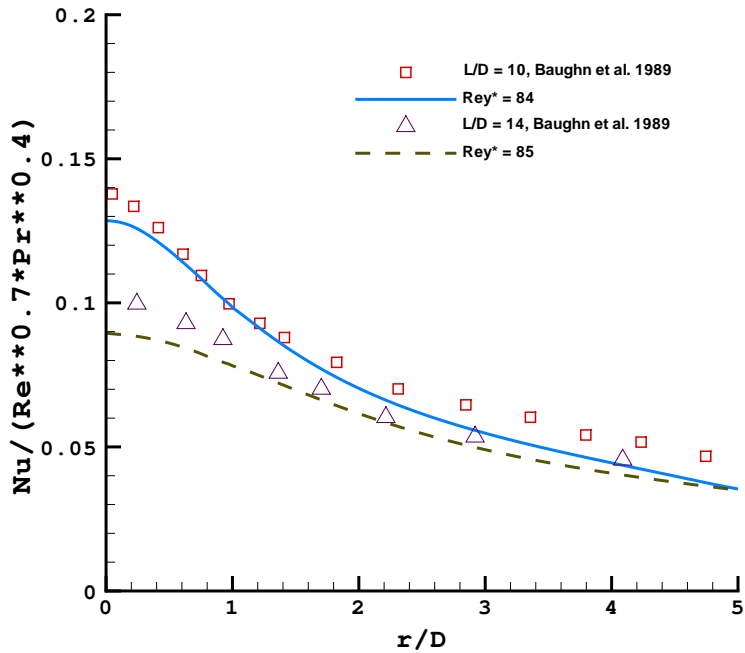}
}
\caption{Local Nusselt number for the impinging jet. Comparison computational solutions against  experimental data for $Re = 23000$ and $L/D = 10; 14$.}
\label{Fig:nusselt10}
\end{figure}

In Fig.\ref{Fig:nusselt10} the distribution of the local Nusselt number is shown for $Re = 23000$. It is given a comparison between the computational results and experimental data for $L=10D$ and $L=14D$. In these examples the wall is located far enough from the nozzle, and in the computations the overprediction of the heat flux nearby the axis of symmetry is not observed. 

\begin{flushleft}
\bfseries 5 Conclusion
\end{flushleft}
The wall functions are formulated as boundary conditions of Robin--type and represented in a differential form. These wall functions take into account source terms. The wall functions are obtained in a compact easy-to-implement analytical form and they do not include any adjustable parameters. The mesh distribution inside the computational domain can be chosen independently on the location of the intermediate boundary. The implementation of the wall functions is robust due to their simultaneous formulation for both a functions and its normal derivative. The Robin--type wall functions are written in a universal formulation applicable to all dependent variables but $\epsilon$ including the kinetic turbulent energy and normal velocity. General approaches to implementing the Robin--type wall functions to finite--volume and finite--difference approximations are suggested. 

On the base of the generalized wall functions implemented to the $k-\epsilon$ model the axisymmetrical impinging jet is investigated. The computational results show a reasonable correspondence to the experimental data and weak dependence of the solution on the distance from a wall where the boundary conditions are set.

Further research can be devoted to application of the wall functions to separated flows and extension of them to LES.

\clearpage

\end{document}